# DISCUSSION OF: TREELETS—AN ADAPTIVE MULTI-SCALE BASIS FOR SPARSE UNORDERED DATA

BY ROBERT TIBSHIRANI

*Stanford University*

This is a very interesting paper on an important topic—the problem of extracting features in an unsupervised way from a dataset. There is growing evidence that unsupervised feature extraction can provide an effective set of features for supervised learning: see, for example, the interesting recent work on learning algorithms for Boltzmann machines [Hinton, Osindero and Teh (2006)].

The ideas in this paper are exciting—treelets are a neat construction that combine clustering and wavelets, and are simple enough to be theoretically tractible. The connection to the latent variable model is also interesting: this kind of model is also the basis of supervised principal components, a method that I co-developed recently [Bair et al. (2006)] for regression and survival analysis in the $p > N$ setting.

I have no practical experience with treelets, so my remaining comments will be brief and mostly in the form of questions for the authors. A much simpler approach to this problem would be to hierarchically cluster the predictors, and then take the average at every internal node of the dendrogram. Let's call this the "simple averaging" method. As noted by the authors, this has already been proposed in the literature, for example, in the "Tree-harvesting" procedure. In this approach we keep all of the original predictors and all of the internal node averages and so end up with an over-complete basis of $2p$ basis functions.

How are treelets different from simple averaging? Treelets do an orthogonalization after each node merge, but does this change the clustering in a material way? What advantage is there to the orthogonal basis delivered by treelets? After all, it looks like the resulting linear combinations of variables are not uncorrelated. Does the simple averaging method perform as well as treelets in the kind of examples of the paper? Do the authors' theorems

---

apply to the simple averaging method as well, or are treelets uniquely good in their estimation of the components of a latent variable model?

The contrast between treelets and simple averaging is analogous to the contrast between wavelets and basis pursuit [Chen, Donoho and Saunders (1998)]. The former is an orthogonal basis while the latter is over-complete; when fitting is done with an $L_1$ (lasso) penalty, the over complete basis, can provide a very good predictive model.

One small point—hierarchical clustering is usually done with average linkage between pairs of predictors. A variation, commonly used in genomics and sometimes called Eisen clustering (since it's implemented in Eisen's Cluster program), uses instead the distance (or correlation) between centroids. The Treelet construction looks more like Eisen clustering. The point is that one could apply Eisen clustering, and then simply average the predictors in every internal node.

DEPARTMENTS OF HEALTH RESEARCH & POLICY,
AND STATISTICS
STANFORD UNIVERSITY
STANFORD, CALIFORNIA 94305
USA
E-MAIL: tibs@stanford.edu